\documentclass{emulateapj}
\usepackage{apjfonts}

\newbox\grsign \setbox\grsign=\hbox{$>$} \newdimen\grdimen
\grdimen=\ht\grsign
\newbox\simlessbox \newbox\simgreatbox \newbox\simpropbox
\setbox\simgreatbox=\hbox{\raise.5ex\hbox{$>$}\llap
     {\lower.5ex\hbox{$\sim$}}}\ht1=\grdimen\dp1=0pt
\setbox\simlessbox=\hbox{\raise.5ex\hbox{$<$}\llap
     {\lower.5ex\hbox{$\sim$}}}\ht2=\grdimen\dp2=0pt
\setbox\simpropbox=\hbox{\raise.5ex\hbox{$\propto$}\llap
     {\lower.5ex\hbox{$\sim$}}}\ht2=\grdimen\dp2=0pt

\slugcomment{}

\shorttitle{A white dwarf-brown dwarf binary in Praesepe}
\shortauthors{Casewell et al.}

\begin{document}


\title{WD0837+185: the formation and evolution of an extreme mass ratio white dwarf-brown dwarf binary in Praesepe.}


\author{S.L. Casewell, M.R.Burleigh, G.A.Wynn and R.D. Alexander}
\affil{Department of Physics and Astronomy, University of Leicester, University Road, Leicester LE1 7RH, UK.}

\email{slc25@le.ac.uk}

\author{R. Napiwotzki}
\affil{Science \& Technology Research Institute, University of Hertfordshire, College Lane, Hatfield, AL10 9AB, UK.}

\author{K.A. Lawrie}
\affil{Department of Physics and Astronomy, University of Leicester, University Road, Leicester LE1 7RH, UK.}

\author{P.D. Dobbie}
\affil{School of Mathematics and Physics, Univeristy of Tasmania, Hobart, Tasmania 7001, Australia.}

\author{R.F. Jameson}
\affil{Department of Physics and Astronomy, University of Leicester, University Road, Leicester LE1 7RH, UK.}

\and

\author{S.T. Hodgkin}
\affil{Institute of Astronomy, Madingley Road, Cambridge, CB3 0HA, UK.}



\begin{abstract}
There is a striking and unexplained dearth of brown dwarf companions in close
orbits ($<$ 3AU) around stars more massive than the Sun, in stark contrast to
the frequency of stellar and planetary companions. Although
rare and relatively short-lived, these systems leave detectable evolutionary
end points in the form of white dwarf - brown dwarf binaries and these
remnants can offer unique insights into the births and deaths of their parent
systems. We present the discovery of a close (orbital separation $\sim$ 0.006
AU) substellar companion to a massive white dwarf member of the Praesepe star
cluster. Using the cluster age and the mass of the white
dwarf we constrain the mass of the white dwarf progenitor star to lie in the
range 3.5 - 3.7 M$_{\odot}$ (B9). The high mass of the white dwarf means
the substellar companion must have been engulfed by the B star's envelope
while it was on the late asymptotic giant branch (AGB). Hence, the initial
separation of the system was $\sim$ 2 AU, with common envelope evolution
reducing the separation to its current value. The initial and final orbital
separations allow us to constrain the combination of the common envelope 
efficiency ($\alpha$) and binding energy parameters ($\lambda$) for the AGB
star to $\alpha\lambda \sim$ 3. We examine the various formation scenarios 
and conclude that the substellar object was most likely to have been 
captured by the white dwarf progenitor early in the life of the cluster, rather than forming in situ.
\end{abstract}





\section{Introduction}

There is a known dearth of brown dwarf companions to solar-type stars with orbital periods $<$5 years (equivalent to orbital separations $<$3AU) when compared with lower mass planetary companions or more massive stellar companions \citep{grether06}. There is also some evidence that this paucity of objects may extend to much larger separations \citep*{mccarthy04}.   The reason for the lack of brown dwarf companions at these separations is unknown, but it is likely related to the formation mechanisms involved.

The difficulties in identifying brown dwarfs with early type-companions mean the most
extreme examples of these binary systems are found in a highly evolved form:
white dwarf - brown dwarf binaries.   However, detached brown dwarf and very
low-mass stellar companions
to white dwarfs are rare; the fraction of L-type secondaries to white dwarfs
is just 0.4$\pm$0.3\% \citep{steele11}. Proper motion surveys and searches for
IR excesses have so far found only a handful of confirmed examples
(\citealt{becklin88, farihi04, maxted06, steele07, steele09, burleigh11,
  dayjones11, debes11}), none of which have a reliably determined age
independent of the white dwarf parameters.  Only in 2 systems,  WD0137-349B
(L8, 0.053M$_{\odot}$; \citealt{maxted06}) and GD1400B (L6-L7,
0.07-0.08M$_{\odot}$; \citealt{farihi04,burleigh11}) (P$_{\rm orb}$ = 116 minutes and 9.8 hours respectively) is the brown dwarf known to have survived a phase of common envelope (CE) evolution. This phase of binary star evolution involves the brown dwarf being engulfed by, and immersed in, the expanding atmosphere of the white dwarf progenitor as it evolves away from the main sequence \citep[see e.g.][]{davis12}. 
In this letter we present the discovery a new white dwarf - substellar binary system in the Prasespe open star cluster. We show how the cluster age along with the mass and cooling age of the white dwarf can be used to place two independent limits on the mass of the white dwarf progenitor star and, additionally, to constrain the initial orbital radius of the brown dwarf. We use this information to examine the physics of common envelope evolution and to test formation models for the original system.

\section{Observations}
\begin{figure}

\includegraphics[angle=270, scale=0.3]{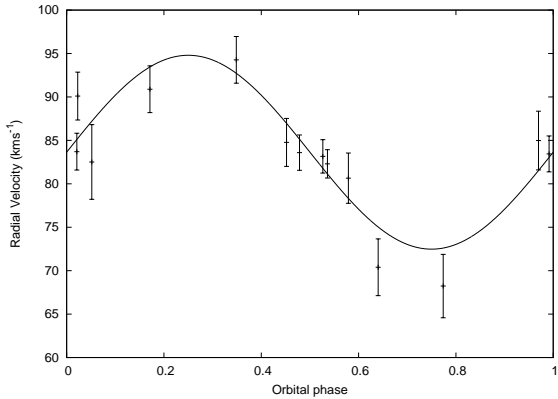}
\caption{Radial velocity phase diagram: The radial velocity data folded on the
  most likely period of 4.2 hours. v$_{\star}$sin i= 11.31$\pm$1.55
  kms$^{-1}$.
\label{fig1}}
\end{figure}

We originally obtained high resolution optical spectra of the 0.798$\pm$0.006
M$_{\odot}$ white dwarf WD0837+185 to confirm its membership of the 625$\pm$50
Myr old Praesepe open star cluster and discovered that its radial velocity was
varying \citep{casewell09}, but that it was not a double lined spectroscopic
binary.  Subsequently, 22 follow-up observations were acquired  between 2008/02/07 and 2008/03/11 from the Very Large Telescope's Ultraviolet Echelle Spectrograph. These data were obtained with exposure times of 20 min in thin cirrus, with seeing between 0.5 and 2.0".  Each pair of consecutive datasets (obtained $\sim$ 1 minute apart) were coadded to increase the S/N. The data were obtained with the same grating settings, reduced and analysed as in \cite{casewell09}. 
These new measurements verified the variation with a best fitting period of 4.2 hours, confirming the presence of an unseen companion. The velocity semi-amplitude calculated from the system parameters is, K = v$_{\star}$ $\sin i =$ 11.31$\pm$1.55 km s$^{-1}$ (Figure \ref{fig1}), giving a minimum mass for the companion M~$\sin i~ \approx$ ~ 25 M$_{\rm Jup}$.
We also obtained photometry from the United
Kingdom Infrared Telescope (UKIRT) using the UKIRT Fast Track Imager in the
$J$, $H$, and $K$, bands. These data had exposure times of 3000s in the $K$,
1200s in the $H$ and 600s in the $J$ band using a 5 point jitter pattern. The data were reduced using the \textsc{starlink} package \textsc{orac-dr}, and the photometry performed using \textsc{iraf}. We also observed WD0837+185 with the $Spitzer$ space telescope using IRAC in the [3.5] and [4.5] micron bands (Cycle 7,Programme ID:57771,  PI: Casewell). These data were reduced using the \textsc{mopex} pipeline, and the aperture photometry performed using \textsc{apex} before the pixel phase, array location and aperture corrections were applied. 

\section{Results}

\begin{figure}
\includegraphics[angle=270,scale=0.3]{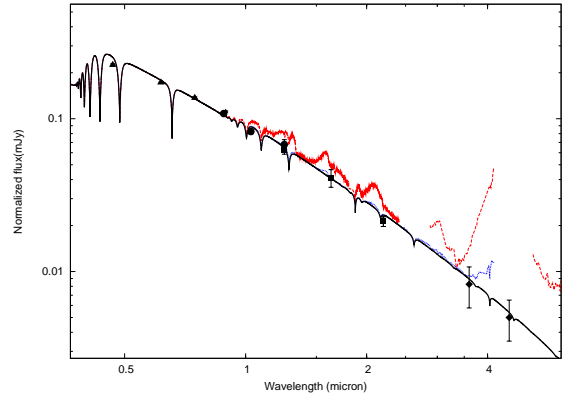}
\caption{Photometry and models of WD0837+185: SDSS $ugriz$ (triangles), UKIRT $ZYJHK$ (boxes) and $Spitzer$ [3.6], [4.5] (diamonds) magnitudes
shown with a DA white dwarf model spectrum (T$_{\rm eff}$=15000 K, log
g=8.3). 3$\sigma$ errorbars are shown on the $Spitzer$ datapoints. A DA+T5 composite spectrum is also shown as a dashed red line, and a
DA+T8 spectrum as the dotted blue line.
The T dwarf spectra are real data and are the objects 2MASSJ05591914-1404488 \citep{cushing06}
and 2MASSJ04151954-0935066 \citep{saumon07}, but the spectra are not continuous as they are $M$
and $L$ band spectra. There are gaps between 4.1 and 5.2 $\mu$m. See the electronic edition of the Journal for a
color version of this figure.\label{fig2}}

\end{figure}

A comparison of optical photometry from the Sloan Digital Sky Survey (SDSS),
near infrared photometry obtained with UKIRT, the UKIRT Infrared Deep Sky
Survey, and mid-infrared images from $Spitzer$ with a pure Hydrogen atmosphere
white dwarf model for WD0837+185 (T$_{\rm eff}=15000$ K, log g$=8.3$;
\citealt{casewell09}), and the Hydrogen model combined with observed spectra
of a T5, and a T8 dwarf showed that no excess emission, indicative of a
companion, is seen. However, if we increase the errors to the 3$\sigma$ level there is a possibility of
an excess in the [3.6] and [4.5] micron wavebands if the white dwarf has a T8
or cooler companion. (Figure \ref{fig2}). 
WD0837+185 has a luminosity, proper motion, radial velocity, cooling age and gravitational
redshift consistent with being a member of Praesepe \citep{casewell09}.  If
the unseen companion were another white dwarf, this would increase the total
luminosity and make it inconsistent with Praesepe membership. Even a high
mass ($\sim 1.38$M$_\odot$) and therefore small radius white dwarf with a cooling age $\sim$ the cluster age (giving an effective temperature of 25000K) would be detected in the
optical photometry \citep{holberg06, tremblay11}. Moreover, if the secondary
were another white dwarf or even a neutron star the radial velocity solution would require an extremely
low inclination orbit ($\sim$1-4 degrees; Probability $<$
2.4$\times$10$^{-03}$) to hide its influence. Hence, we rule out the
possibility that WD0837+185B is another white dwarf or a neutron star.  WD0837+185 is not
coincident with an X-ray source in the ROSAT all sky survey, eliminating the
possibility the system contains an accreting black hole. The probability of the system containing a non-accreting black hole is 6$\times$10$^{-6}$, as the inclination would need to be extremely low ($\sim$0.2 degrees) to match radial velocity measurements.

The photometry in the [4.5] micron band gives the maximum possible mass of a
non-degenerate secondary as 30M$_{\rm Jup}$ \citep{baraffe03}.  Combined with
the minimum mass from the radial velocity solution (M $\sin i \approx$ 25
M$_{\rm Jup}$) this strongly suggests that  the companion is substellar and is
likely to be a methane atmosphere T-type brown dwarf (T8 or later) with an effective temperature of $\approx$900-1100 K \citep{baraffe03} (Figure \ref{fig2}).

\begin{figure}
\includegraphics[angle=0,scale=0.5]{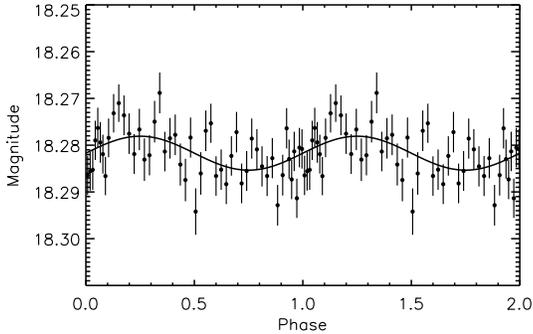}
\caption{Light curve from March 2009 binned by a factor of 7 folded on the orbital period of 4.2 hours. The peak to peak variation is at 0.74\%.\label{fig3}}
\end{figure}

We have also obtained V band photometry obtained every 30s over a total of 5
hours from the Isaac Newton Telescope on La Palma in March 2009 to investigate
the possibility of an eclipse in the system. The data show variation at the
0.74\% level (peak to peak) on the orbital timescale (Figure \ref{fig3}). This
may be indicative of irradiation by the high ultraviolet flux of the white
dwarf, possibly leading to substantial temperature differences between the
``day'' and ``night'' hemispheres of the T dwarf atmosphere as is seen in
WD0137-349 \citep{maxted06}. No eclipse is seen, but the probability of
eclipse in this system is only $\sim$9\% and any eclipse would only last
$\sim$3.5 minutes \citep{faedi11}, which is difficult to detect with our sampling frequency (30s exposure + 30s readout).
The main-sequence lifetime of the white dwarf progenitor is constrained by the
difference between the cooling age of WD0837+185 (313$\pm$5 Myr, determined
from the white dwarf temperature, gravity and appropriate cooling model; \citealt{fontaine01}) and the age of the Praesepe cluster (625$\pm$50Myr), which limits the progenitor mass to 3.48$\pm$0.23M$_{\odot}$ equivalent to a spectral type  B9. Intriguingly, the orbit of WD0837+185B (orbital separation $\sim$0.006 AU $\sim$1.24R$_{\odot}$) is well within the main sequence radius of a B9 star (R $\sim$3R$_{\odot}$; \citealt{marigo08}). WD0837+185B is unlikely to have been captured into this current orbit as the star-crossing timescale in Praesepe is a few Myr and the low stellar density (20 stars\/pc$^{-2}$; \citealt{kraus07}) makes the typical close encounter ($\ll$ 1AU) time-scale much longer than the age of the Universe. The system must therefore have gone through a phase of CE evolution. This brief evolutionary phase is a key stage in the formation of short period binary systems and occurs when the lower mass companion becomes engulfed by the nuclear-evolution driven expansion of the primary star's envelope. The binary orbit then decays rapidly as drag forces unbind the primary's envelope at the expense of orbital energy. 
		
The mass of WD0837+185 ($\sim$0.8M$_{\odot}$) is at the very upper end of core masses attainable within the errors on the progenitor star mass. Models relating initial stellar mass to core mass (taken to be the white dwarf mass) at first thermal pulse \citep{karakas02} give a lower limit of $\sim$ 3.6M$_{\odot}$ for the progenitor mass, while the cooling age-cluster age argument above provides an upper limit of $\sim$ 3.7M$_{\odot}$. The very limited overlap between these independent mass estimates show that the progenitor star engulfed the brown dwarf very late in its evolution on the AGB. The radius of the progenitor at this point, and hence the orbital separation of WD0837+185B at the onset of the CE phase, would have been $\sim$~2AU \citep{ventura09}. This estimate places the orbital separation of the original system well within the region where there is an observed dearth of brown dwarf companions to solar-type stars. 

Treatments of CE evolution are usually parameterized in terms of the efficiency with which orbital energy is transferred to the envelope of the primary star ($\alpha$) and a parameter governing the binding energy of the primary's envelope ($\lambda$)(see e.g.\ \citealt{davis12,xu10}), the latter being defined by
\begin{equation}
E_{\rm bind}=-\frac{GM_{1}M_{\rm env}}{\lambda R_{1}},
\end{equation}
where $E_{\rm bind}$ is the binding energy, M$_{1}$ is the progenitor (primary) mass, M$_{\rm env}$ is progenitor envelope mass and R$_{1}$ is the progenitor's radius. Many population synthesis calculations treat $\lambda$ as a constant, but its value is poorly constrained. Treatments differ in the inclusion (or not) of the internal energy of the stellar matter. We are able to place explicit limits on $\lambda$ because we know that the CE phase began when the WD0837+185 progenitor was on the late AGB, which fixes the original orbital radius as well as the core and the envelope mass of the progenitor. Knowledge of the initial (a$_{i} \sim$ R$_{1}$) and final (a$_{f}$) orbital separations allows the combination $\alpha \lambda$ to be extracted from the energy balance equation
\begin{equation}
\alpha \lambda \left(\frac{G M_{core} M_{2}}{2a_{f}}-\frac{GM_{1}M_{2}}{2a_{i}}\right)=\frac{GM_{1}M_{env}}{R_{1}},
\end{equation}
where M$_{2}$ is the brown dwarf mass and M$_{\rm core}$ (= M$_{1}$ - M$_{\rm env}$) is the progenitor core mass. For a progenitor mass 3.6 M$_{\odot}$ and a companion mass of 30M$_{\rm Jup}$ the combination $\alpha \lambda$ $\sim$3 is required to place the brown dwarf in its current orbit from an initial orbit $\sim$2AU, assuming no significant mass loss by the progenitor at the point of contact, (any mass loss would cause a proportionally lower estimate of $\alpha \lambda$). In the case of maximum efficiency ($\alpha$ =1), $\lambda \sim$3 is very low for a highly evolved 3.6 M$_{\odot}$ star if the internal energy of the stellar material is taken into account when calculating $\lambda$, but is in reasonable agreement with calculations including only gravitational binding energy \citep{xu10}.  Highly evolved, late AGB primaries of this mass are precisely those predicted to have the highest values of $\lambda$ when including the contribution of the internal energy of the stellar matter, and a low $\lambda$ for these objects is a strong hint that this parameter is of order unity for all primaries. 

\section{Interpretation}
WD0837+185B may have formed (at its original orbital separation) in one of two ways: in a manner similar to Solar System and extra-solar giant planets, or as an extreme mass-ratio binary star. The planetary channel assumes formation in a disk around the newly-formed star, and consists of two options: core accretion or gravitational instability.  Core accretion is not a promising formation mechanism here: tidal torques from the planet strongly suppress gas accretion from the disk on to the planet for masses $\gtrsim$ 5M$_{\rm Jup}$ \citep{dangelo02}, and it is unlikely that objects as massive as as $\sim$20M$_{\rm Jup}$ can ever form in this manner.  Moreover, the growth time-scale for such a massive object is at least comparable to the typical lifetime of protoplanetary disks around B stars (which are estimated to be Myr or less; \citealt{hillenbrand92}).

Disk fragmentation via gravitational instability can result in the formation of much more massive objects, and is a plausible scenario for brown dwarf formation \citep[e.g.,][]{stamatellos07}, but carries the caveat
that protostellar disks are gravitationally unstable only at large radii.  For
a 3.5 M$_{\odot}$ star a self-luminous disk is only unstable at radii $\geq$70
AU  \citep{lm05,rafikov09}, and this critical radius increases when
irradiation from the star is taken into account.  The question then becomes
whether a brown dwarf can migrate inwards to 1-2AU from its formation radius
at $\sim$100 AU.  Dynamical ``hardening" of the system via repeated encounters
with other cluster stars is extremely unlikely, as repeated such interactions
are more likely to disrupt the system than shrink its orbit; also, the low
stellar density in Praesepe essentially rules out this mechanism.  Inward
migration via gravitational interactions with other brown dwarf- or
planetary-mass objects (so-called planet-planet scattering) is possible, but
again unlikely: the process is chaotic, but simulations find that inward
migration via this mechanism is usually modest \citep[e.g.,][]{raymond08}; the
probability of being scattered from $\sim$100 AU to $\sim$1-2 AU is very
small.  The final possibility is gas-driven migration before the dispersal of
the protostellar disk.  Giant planets typically migrate in the Type II regime.
Although the discovery of two hot Jupiters around stellar members of Praesepe by
\citet{quinn} shows that planet formation and migration in the Type II regime
has occurred in this open cluster, it is unlikely that objects as massive as
$\sim$25 M$_{\rm Jup}$ can ever form in this manner. For an object of 25-30 M$_{\rm Jup}$ Type II migration is strongly suppressed and proceeds on a time-scale much longer than the lifetime of the gas disk \citep{syer95}, but recent simulations have found that migration in gravitationally unstable disks can in fact be very rapid \citep{cha11,baruteau11}.  It is unclear, however, whether it is possible to halt the rapid migration of 20-30 M$_{\rm Jup}$ objects at $\sim$1 AU, and prevent them falling all the way on to the central star.  This mechanism cannot be ruled out without further investigation, but we do not consider it to be the most likely formation channel for WD0837+185B.

Binary stars form from the fragmentation of star-forming molecular cores, but are affected by the cluster dynamics as they evolve. Extreme mass-ratio binaries in close orbits such as this one ($q\simeq0.01$) are very rare.  Forming such systems in a similar manner to binary stars is challenging, because circumstellar material preferentially accretes on to the secondary, increasing its mass and driving the binary mass-ratio towards $q = 1$ \citep{arty83}. Many theoretical studies have investigated how dynamical interactions may harden initially wide binaries, or destroy them, but the fundamental physics of binary formation remains poorly understood \citep[see, e.g.,][and references therein]{goodwin07}.  
            
The one remaining formation theory left for consideration is thus dynamical capture. Recent numerical simulations \citep{bate11} show that 
in a N$\sim$100-500 cluster a few objects are formed this way, though due to the low number statistics it is not clear how frequently 
extreme mass-ratio objects are captured into $\sim$ AU orbits.   Praesepe has more members than in this example (N$\sim$1000;
\citealt{kraus07}), and at its current age, is likely to have undergone dynamical evolution, decreasing the cluster population 
by as many as half, and preferentially ejecting the lowest mass cluster members \citep{delafuente02}.  The most plausible 
formation scenario for WD0837+185B is therefore likely to be dynamical capture where the brown dwarf has been inserted 
into the binary after its formation. This is likely to be a common formation scenario for those relatively rare oases 
in the brown dwarf desert: high mass ratio main sequence plus brown dwarf pairs found by radial velocity surveys with 
separations of a few AU \citep[e.g.,][]{omiya}.

\section{Conclusions}
We confirm that WD0837+185 is a radial velocity variable object and conclude
from optical, near- and mid-IR photometry
that the probable companion is a 25-30M$_{\rm Jup}$ late T dwarf. Optical
photometry also tentatively indiactes that the white dwarf is irradiating its
substellar companion, although no eclipse is seen in the data.

Using the cluster age and the mass of the white
dwarf we constrain the mass of the white dwarf progenitor star to lie in the
range 3.5 - 3.7 M$_{\odot}$ (B9). The high mass of the white dwarf means
the substellar companion must have been engulfed by the B star's envelope
while it was on the late AGB. Hence, the initial
separation of the system was $\sim$2 AU, with common envelope evolution
reducing the separation to its current value. The initial and final orbital
separations allow us to constrain the combination of the common envelope 
efficiency ($\alpha$) and binding energy parameters ($\lambda$) for the AGB
star to $\alpha\lambda \sim$3. We examine the various formation scenarios 
and conclude that the substellar object was most likely to have been 
captured by the white dwarf progenitor early in the life of the cluster, rather than forming in situ.

\acknowledgments
We thank Mike Cushing and Didier Saumon for providing the comparison spectra of the T5 (2MASSJ05591914-1404488) and T8 (2MASSJ04151954-0935066) white dwarfs.
Based on observations made with ESO telescopes at the La Silla Paranal
Observatory under programme ID 080.D-0456(A) and the spectra used in this work
are available in the ESO science archive. This work is also based on
observations made with the INT operated on the island of La Palma by the Isaac
Newton Group in the Spanish Observatorio del Roque de los Muchachos of the
Instituto de Astrofisica de Canarias.  Observations were also made with the
Spitzer Space Telescope, which is operated by the Jet Propulsion Laboratory,
California Institute of Technology under a contract with the National
Aeronautics and Space Administration. RDA acknowledges support from the
Science \& Technology Facilities Council (STFC) through an Advanced Fellowship
(ST/G00711X/1). SLC acknowledges support from the Univeristy of Leicester.  Theoretical Astrophysics and Observational Astronomy in Leicester is supported by an STFC Rolling Grant.


\begin{thebibliography}{}
\bibitem[Artymowicz(1983)]{arty83} Artymowicz, P. 1983, Acta Astron., 33, 223

\bibitem[Baraffe et al.(2003)]{baraffe03} Baraffe, I., Chabrier, G., Barman, T.S., Allard, F., \& Hauschildt, P.H. 2003, \aap, 402, 701

\bibitem[Baruteau, Meru \& Paardekooper(2011)]{baruteau11} Baruteau, C., Meru, F., \& Paardekooper, S.-J. 2011, \apj, 416, 1971

\bibitem[Bate(2011)]{bate11} Bate, M.R. 2011, \mnras, 417, 2036

\bibitem[Becklin \& Zuckerman(1998)]{becklin88} Becklin, E.E., \& Zuckerman, B. 1998, \nat, 336, 656

\bibitem[Burleigh et al,(2011)]{burleigh11} Burleigh M.R., et al, 2011, in
  American Institute of Physics Conference Series 1331, Planetary systems
  beyond the main sequence ed. Schuh, Dreschel \& Heber, 289

\bibitem[Casewell et al.(2009)]{casewell09} Casewell, S.L., Dobbie, P.D., Napiwotzki, R., Burleigh, M.R., Barstow, M.A., \& Jameson, R.F. 2009, \apj, 395, 1795

\bibitem[Cha \& Nayakshin(2011)]{cha11} Cha, S.-H., \& Nayakshin, S. 2011, \mnras, 415, 3319

\bibitem[Cushing et al.,(2006)]{cushing06} Cushing, M.C. et al. 2006, \apj,
  648, 614

\bibitem[D'Angelo, Henning \& Kley(2002)]{dangelo02} D'Angelo, G., Henning, T., \& Kley, W. 2002, \apj, 385, 647

\bibitem[Davis, Kolb \& Knigge(2012)]{davis12} Davis, P.J., Kolb, U., \& Knigge, C. 2012, \mnras, 419, 287

\bibitem[Day-Jones et al.(2011)]{dayjones11} Day-Jones, A.C. et al., 2011, \mnras, 410, 705

\bibitem[Debes et al.(2011)]{debes11} Debes, J.H., Hoard, D.W., Wachter, S., Leisawitz, D.T., \& Cohen, M. 2011, \apjs, 197, 38

\bibitem[de La Fuente Marcos \& de La Fuente Marcos(2002)]{delafuente02} de La
  Fuente Marcos, R., \& de La Fuente Marcos, C. 2002, in Astronomical Society
  of the Pacific Conference Series 285, Modes of star formation and the origin of field populations ed. Grebel \& Brandner, 170

\bibitem[Faedi et al.(2011)]{faedi11} Faedi, F., West, R.G., Burleigh, M.R., Goad, M.R., \& Hebb, L. 2011, \mnras, 410, 899

\bibitem[Farihi  \& Christopher(2004)]{farihi04} Farihi, J., \& Christopher, M. 2004, \aj, 128, 1864

\bibitem[Fontaine, Brassard, \& Bergeron(2001)]{fontaine01} Fontaine, G., Brassard, P., \& Bergeron, P. 2001, PASP, 113, 409

\bibitem[Goodwin et al.(2007)]{goodwin07} Goodwin, S.P., Kroupa, P., Goodman,
  A., \& Burkert, A. 2007, in Protostars and Planets V ed. Reipurth, Jewitt,
  \& Keil (Tuscon, AZ:University of Arizona press)  133 

\bibitem[Grether \& Lineweaver(2006)]{grether06} Grether, D., \& Lineweaver, C.H. 1982, \apj, 640, 1051

\bibitem[Holberg \& Bergeron(2006)]{holberg06} Holberg, J., \& Bergeron, P.,
  2006, \aj, 132, 1221

\bibitem[Hillenbrand et al.(1992)]{hillenbrand92} Hillenbrand, L.A., Strom, S.E., Vrba, F.J., \& Keene, J. 1992, \apj, 397, 613

\bibitem[Karakas, Lattanzio \& Pols(2002)]{karakas02} Karakas, A.I., Lattanzio, J.C., \& Pols, O.R. 2002, PASA, 19, 515

\bibitem[Kraus \& Hillenbrand(2007)]{kraus07} Kraus, A.L., \& Hillenbrand, L.A. 2007, \aj, 134, 2340


\bibitem[Marigo et al.(2008)]{marigo08} Marigo, P., Girardi, L., Bressan, A., Groenwegen, M.A.T., Silva, L., \& Granato, G.L. 2008, \aap, 482, 883

\bibitem[Matzner \& Levin(2005)]{lm05} Matzner, C.D, \& Levin, Y. 2005, \apj,
  628, 817

\bibitem[Maxted et al.(2006)]{maxted06} Maxted, P.F.L., Napiwotzki, R., Dobbie, P.D., \& Burleigh, M.R. 2006, \nat, 442, 543

\bibitem[McCarthy \& Zuckerman(2004)]{mccarthy04} McCarthy, C., \& Zuckerman, B. 2004, \aj, 127, 2871

\bibitem[Omiya et al.(2011)]{omiya} Omiya, M., et al., 2011 in 
  American Institute of Physics Conference Series 1331, Planetary systems
  beyond the main sequence ed. Schuh, Dreschel \& Heber, 122 

\bibitem[Quinn et al.(2012)]{quinn} Quinn, S.N. et al., 2012, \apj, 756, L33 

\bibitem[Rafikov(2009)]{rafikov09} Rafikov, R.R. 2009, \apj, 704, 281

\bibitem[Raymond et al.(2008)]{raymond08} Raymond, S.N., Barnes, R., Armitage, P.J., \& Gorelick, N. 2008, \apjl, 687, L107

\bibitem[Saumon et al.,(2007)]{saumon07} Saumon, D., et al. 2007, \apj, 656,
  1136

\bibitem[Stamatellos, Hubber \& Whitworth(2007)]{stamatellos07} Stamatellos, D., Hubber, D.A., \& Whitworth, A.P. 2007, \mnras, 382, L30

\bibitem[Steele et al.(2007)]{steele07} Steele, P.R., Burleigh, M.R., Dobbie, P.D., \& Barstow, M.A. 2007, \mnras, 382, 1804

\bibitem[Steele et al.(2009)]{steele09} Steele, P.R., Burleigh, M.R., Farihi, J., G{\"a}nsicke, B.T., Jameson, R.F., Dobbie, P.D., \& Barstow, M.A. 2009, \aap, 500, 1207

\bibitem[Steele et al.(2011)]{steele11} Steele, P.R., Burleigh, M.R., Dobbie, P.D., Jameson, R.F., Barstow, M.A., \& Satterthwaite, R.P. 2011, \mnras, 416, 2768

\bibitem[Syer \& Clarke(1995)]{syer95} Syer, D., \& Clarke, C.J. 1995, \mnras, 277, 758

\bibitem[Tremblay, Bergeron \& Gianninas(2011)]{tremblay11} Tremblay, P-E.,
  Bergeron, P., \& Gianninas, A. 2011, \apj, 730, 128

\bibitem[Ventura \& Marigo(2009)]{ventura09} Ventura, P., \& Marigo, P. 2009, \mnras, 399, L54

\bibitem[Xu \& Li(2010)]{xu10} Xu, X.-J., \& Li, X.-D. 2010, \apj, 716, 114

\end{thebibliography}
\end{document}